\documentstyle[aps,prd,graphicx,twocolumn]{revtex}
\begin{document}

\def\beq{\begin{equation}}
\def\eeq{\end{equation}}
\def\bea{\begin{eqnarray}}
\def\eea{\end{eqnarray}}
\def\thet{\theta_{\mu}}
\def\Sig{\Sigma_{\mu}}
\def\eps{\epsilon_i}
\def\s{${\bf s }$}
\def\p{${\bf \Theta }$}

\twocolumn[\hsize\textwidth\columnwidth\hsize\csname
@twocolumnfalse\endcsname

\title{Eyes Wide Open - Optimising Cosmological Surveys in a Crowded Market}
\author{Bruce A. Bassett}
\address{Department of Physics, Kyoto University, Kyoto, Japan  \& \\ Institute of Cosmology and Gravitation, 
University of Portsmouth, Portsmouth~PO1~2EG, UK}
\maketitle

\begin{abstract}
Optimising the major next-generation cosmological surveys (such as {\em SNAP, KAOS etc...}) is a key problem given our ignorance of the physics underlying cosmic acceleration and the plethora of surveys planned. We propose a Bayesian design framework which (1) maximises the discrimination power of a survey without assuming any underlying dark energy model, (2) finds the best niche survey geometry given current data and future competing experiments, (3) maximises the cross-section for serendipitous discoveries and (4) can be adapted to answer specific questions (such as `is dark energy dynamical?'). Integrated Parameter Space Optimisation (IPSO) is a design framework that integrates projected parameter errors over an entire dark energy parameter space and then extremises a figure of merit (such as Shannon entropy gain which we show is stable to off-diagonal covariance matrix perturbations) as a function of survey parameters using analytical, grid or MCMC techniques. We discuss examples where the optimisation can be performed analytically. IPSO is thus a general, model-independent and scalable framework that allows us to appropriately use prior information to design the best possible surveys. 
\end{abstract}
\maketitle
\vskip 3ex
]

\section{Introduction}
Our almost total ignorance of the source of cosmic acceleration has provided the dark, damp conditions ideal for the spawning of wild and varied theories. Among many others, cosmic acceleration has been ascribed to a modification of gravity on large scales\cite{grav}, macroscopic quantum effects (e.g. \cite{PR}), condensates \cite{cond1,cond2}, unified dark energy\cite{unified}, a late-time phase transition \cite{ltpt} or a `mundane' scalar field with almost flat potential. Unfortunately perhaps, the least informative possibility - Einstein's greatest blunder; a cosmological constant - is still a good fit to current data\cite{longDE}. 

This zoo of possibilities highlights the profound reordering of our view of cosmology and high-energy physics that will follow from understanding the true nature of dark energy. This exciting prospect has stimulated the proposal of a spectacularly wide variety of dark energy experiments for deployment over the next two decades.

Not surprisingly these experiments currently operate on a mutually competitive basis. As a result there has been little or no consideration given to how to optimally configure these surveys in order to get the best, {\em model-independent} constraints on dark energy models either from each survey alone or in conjunction with the other planned surveys. It is the aim of this paper to begin to address these important issues by presenting a framework which we call Integrated Parameter Space Optimisation (IPSO). IPSO implements survey design that is model-independent, flexible and uses prior information within the framework of Bayesian optimal design. IPSO has been implemented numerically using Markov Chain Monte Carlo (MCMC) methods in \cite{ipsoletter} and in optimising the design of the KAOS/GWFMOS instrument.

A cursory look at the descriptions of most next-generation dark energy experiments would suggest that their aim begins and ends with nailing down the two numbers entering the simple expansion $w(z)=w_0 + w_1 z$, of the dark energy equation of state. The truth is very different. 

In order for dark energy experiments to give us new knowledge about high-energy  physics we need to know a great deal more than just the low-redshift evolution of $w(z)$. Knowing the mass of the dark energy particle (equivalently its Compton wavelength\cite{cond1}), its speed of sound\cite{sound} and its couplings to baryons and dark matter\cite{couple} may all prove crucial. Extracting all this information will require the full range of next-generation experiments and beyond. From this perspective, mutually optimising the next-generation experiments to maximise our knowledge is not only prudent, it is crucial.
 
Current and proposed dark energy surveys fall into one of a number of categories. There are tests sensitive to the background dynamics of the cosmos, such as distance (luminosity or angular-diameter) tests. Primary next-generation experiments in this category are the SNAP satellite \cite{snap} and the KAOS baryon oscillation galaxy survey \cite{kaos,se,blake,linder}. There are also Hubble constant tests (such as KAOS) \cite{kaos,chron} and galaxy and cluster number count surveys either using galaxies (such as DEEP2) \cite{deep2} or clusters detected through X-ray emission or the Sunyaev-Zel'dovich effect, such as ACT, SPT, SZA, DUO, DUET \cite{duet} and AMiBA\cite{amiba}.  

Additional powerful constraints come from the CMB which tests both dark energy dynamics and perturbations \cite{longDE}, while VISTA, LSST\cite{lsst} and pan-STARRS \cite{pan} will provide powerful new arenas to search for CMB-LSS correlations, expected from the acceleration-induced late ISW effect, and weak gravitational lensing. The latter will also be probed by SNAP\cite{snaplens}. The proposed Dark Energy Survey (DES) \cite{des} would simultaneously obtain weak lensing, cluster number counts and supernovae data. Further on, the Square Kilometer Array (SKA) will provide excellent constraints on distance and Hubble constant from weak lensing and the baryon oscillations in the matter power spectrum \cite{chris2}. Then there will be unexpected sources of help. For example,  SKA will map the reionisation of the universe, thereby providing independent constraints on $\tau$, the optical depth to Thompson scattering. Since this is degenerate in the TT spectrum of the CMB with the epoch of acceleration, breaking this degeneracy may significantly improve constraints on dark energy dynamics \cite{longDE}.

However, given that many of the experiments listed above are currently lacking a final science definition, a crucial question is ``What is the optimal survey structure (niche) for each of these experiments given the other experiments?" For example, for redshift surveys which measure some quantity as a function of $z$ (which will be the main focus of this paper), what distribution of redshift bin error bars (or equivalently how much observing time should be spent in each redshift bin) should one aim to achieve to get the best constraints on dark energy? Should one concentrate on a small redshift range covering a large area or should one cover a wide range of redshifts in a narrow, pencil-beam survey? All this has to be addressed despite us being fundamentally ignorant about the properties of the dark energy.

Clearly these are difficult questions. First, one does not know in advance which experiments will actually be funded and secondly it is possible that our understanding of the universe will undergo further shocks and revelations hence and designing surveys that are robust and sensitive to the unexpected is desirable. 

This paper presents a framework for optimising any survey (not just a dark energy redshift survey) in a manner which is flexible and easy to adjust if new `competing' experiments are introduced, thereby allowing a clear niche to be found, a niche which can be touted to funding agencies, rather than relying on minimal improvements to errors on $w_0,w_1$. Further, it does not assume a model for the dark energy and it can be automatically adjusted to allow for optimal answering of specific questions. 

Currently, optimisation of surveys in the context of dark energy is at a rather basic stage, primarily perhaps because of our ignorance of dark energy, as discussed earlier. By contrast, optimisation for CMB experiments is rather well understood (see e.g.\cite{cmbopti,pol}). Previous dark energy survey analysis have typically fallen into two categories:  (1) optimisation of survey properties  \cite{HT,SS,y} to best estimate parameters assuming constant $w$ (often $w=-1$) and (2) comparison of error estimates for parameters in a small number (less than six) of specific dark energy models \cite{se,WL,LH}. 

But, except for the proposal in \cite{HT}, no precise criterion for selecting one survey geometry over another has been given. More importantly perhaps, none of these analyses were model-independent and hence were limited for two reasons: first they all selected an underlying dark energy model \footnote{They were not covariantly formulated in the space of dark energy models.} and second, the underlying models chosen typically belonged to a very limited dark energy parameter space, ${\bf \Theta}$, such as $w=constant$ or  $w_0,w_1$ as appearing in the simple expansion $w(z) = w_0 + w_1 z$.

Unfortunately, such an approach induces significant bias when used with real data\cite{params} and does not sufficiently allow for our ignorance of the underlying model. Our proposal (IPSO) is the natural generalisation of this general ethos to a Bayesian optimal design context which allows for model-independence, a specific criterion for optimisation and inclusion of prior and competing survey data. 

In this paper we follow the Einstein summation convention, so repeated indices are summed over.\footnote{So $A_{\mu\nu} B^{\mu\nu} = \sum_{\mu,\nu} A_{\mu\nu} B^{\mu\nu}$.} We interchangebly use index and index-free forms of vectors and matrices, e.g. both ${\bf F}$ and $F_{\mu\nu}$ refer to the expected Fisher matrix. 

Section (\ref{basic}) introduces the framework, section (\ref{fom}) discusses various Figures of Merit (FoM) while section (\ref{optim}) discusses issues related to the actual survey optimisation. Section (\ref{egs}) discusses simple examples of optimisation and the issues one faces in full implementations. 
\begin{figure}
\begin{center}
\includegraphics[width=65mm]{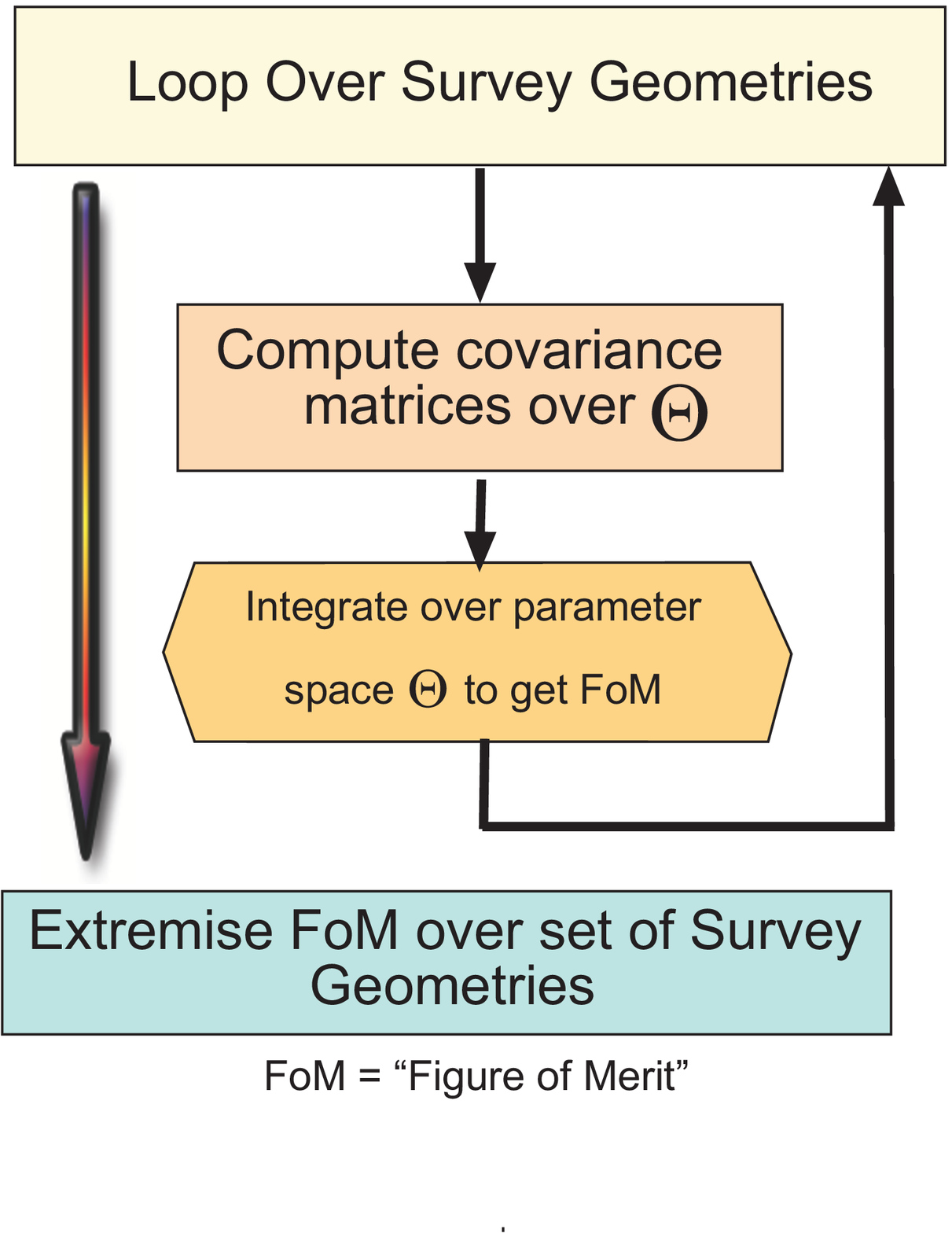} \\
\caption[survey]{Flow-chart for integrated parameter space optimisation (IPSO). For each survey geometry in a denumerable set (either discrete or continuous) some function of the projected parameter covariance matrix is integrated over the dark-energy parameter space ${\bf \Theta}$ to yield a Figure of Merit (FoM, a positive real number). Optimisation proceeds by selecting the candidate survey geometry with the minimum or maximum FoM, depending on the precise FoM used; see section (\ref{fom}).}
\label{flowchat}
\end{center}
\end{figure}
\section{The basic framework for IPSO}\label{basic}

The flow-chart for the Integrated Parameter Space Optimisation (IPSO) framework is simple, endowing it as a result with considerable flexibilty. Consider a set of allowed survey geometries, indexed by \s. As we will discuss below, they can depend continously on survey parameters (such as survey volume or area) or can be discontinuous proposals for different survey geometries or even survey types. 

For example, one geometry may include a completely different component (such as the weak lensing survey added to SNAP, follow-up observations for SNIa in a lensing survey or adding a high-redshift component to a number counts survey). For each survey geometry, \s, we compute an appropriate {\em Figure of Merit (FoM)} - also known as the {\em utility} in Bayesian evidence design, {\em risk} or {\em fitness}  - and optimisation then simply proceeds by selecting the survey geometry which extremises (minimising or maximising where appropriate) the FoM \footnote{Recent applications of FoM optimisation to problems in experiement design in astronomy and astrophysics can be found in \cite{examples}.}. 

The key to successful optimisation clearly lies in the criteria used to construct the FoM. 
Here we propose an FoM suited to our current (lack of) knowledge of dark energy, and to our goals: 
to maximise discrimination power in terms of dark energy models and fundamental physics. Hence, our IPSO FoM is chosen to be an integral of some function, $I$, of the $1\sigma$ covariances over the dark energy parameter space, ${\bf \Theta}$, chosen by the user: 
\beq
FoM(s) = \int_{\Theta} I(s,\thet) d\thet\,.
\label{basicdef}
\eeq
$I(s,\thet)$ is a scalar which in general will depend on the survey geometry, the prior information available and position in ${\bf \Theta}$. We will discuss in the next section suitable choices for $I$.  

By integrating over the whole range of possible dark energy models we achieve model-independence. 
In general ${\bf \Theta}$ will be an n-dimensional space spanned by $n$ dark energy parameters ${\theta_1,...\theta_n}$. A standard example would be the space spanned by $\vec{\theta} = (w_0,w_1,...)$, the parameters entering a description of the dark energy equation of state $w(z)$, but the parameter space could parametrise any quantity related to dark energy (scale factor, Hubble constant, distances, dark energy density, speed of sound, Compton wavelength etc...). 

Our choices for $I$ will be based around the computation,
{\em at each point} in ${\bf \Theta}$, of the expected $1\sigma$ error bars for each of the $n$ parameters spanning ${\bf \Theta}$\footnote{If prior information is to be used, the Fisher matrix following from the prior information (the prior precision matrix) is computed too.}. This can either be done using Fisher matrix techniques or e.g. by direct Monte Carlo Markov Chain (MCMC) simulation. The key point is that the simulated error bars used to compute the likelihood at each grid point will depend on the survey configuration, \s, which is where the power to optimise the survey arises, whereas the prior precision matrix, denoted ${\bf P}$, will be independent of the survey of course. 

Optimisation then consists of two parts. 
\begin{itemize}
\item For each survey configuration, \s, compute $FoM(s)$, given by eq. (\ref{basicdef}), a real number. 
\item Extremise the $FoM(s)$ over the set of survey geometries using analytical, grid or Markov-Chain Monte Carlo (MCMC) methods 
\end{itemize}
We now discuss each of these aspects in turn.   

\section{FoM for a given survey geometry}\label{fom}

There are a number of ways of defining an appropriate Figure of Merit (FoM), which for us reduces to a choice of $I(s,\thet)$ in eq. (\ref{basicdef}). Within the context of Bayesian optimal design this is a well-known problem with many different proposals having been made, depending on what the aim of the experiment is \cite{L72}. Here we consider three choices. 

The first effectively integrates the sum of the components of the $1\sigma$ error-vector over ${\bf \Theta}$ and is in simplest form reduces to A-Optimality in Bayesian design. The second integrates the volume of the $1\sigma$ error ellipse over ${\bf \Theta}$ while the third integrates the logarithm of the error ellipse volume and in simplest form reduces to D-Optimality, namely the maximisation of the Shannon information in going from prior to posterior. All of of these proposals can either include or exclude prior information as desired to yield Bayesian or non-Bayesian optimal solutions respectively.

Before discussing the details we briefly remind the reader of our notation. The dark energy parameters are denoted $\thet$, spanning the $n$-dimensional space ${\bf \Theta}$. The various survey configurations being considered are denoted by ${\bf s}$. The Fisher matrix for the $\thet$ resulting from all prior cosmological information (the prior precision matrix) is denoted ${\bf P}$. For example, it may be derived from CMB, LSS and SNIa data, as in \cite{longDE}. Greek indices ($\mu,\nu...$) label coordinates in ${\bf \Theta}$; roman indices ($i,j...$) label redshift bins.

\subsection{Fisher matrix integration \& A-optimality}

Consider the covariance matrix, labelled $C_{\mu\nu}$, over ${\bf \Theta}$. The $n$-th entry on the diagonal gives the variance in our knowledge of $\theta_n$ while the off-diagonal terms provide the covariances between the parameters. For nonlinear problems $C_{\mu\nu} = C_{\mu\nu}(\thet)$, the error bars in general depend on where one is in ${\bf \Theta}$, the dark energy parameter space\footnote{For example, errors on both $w_0, w_1$ roughly double if the underlying model is a cosmological constant rather than a $w_0=-2/3, w_1=0$ model\cite{se}.}

We can then define a very general FoM to be maximised via eq. (\ref{basicdef}) with the choice:
\bea
I(s,\thet) &&= (C^{-1}(s,\thet))_{\mu\nu} W^{\mu\nu}(\thet)\nonumber\\
&&\simeq F_{\mu\nu} W^{\mu\nu} \equiv \sum_{\mu,\nu} F_{\mu\nu} W^{\mu\nu}\nonumber\\
&&= \mbox{tr}\{{\bf F}~{\bf W}\}
\label{int1}
\eea
where $\mbox{tr}\{{\bf A}\}=\mbox{Tr}\{\bf A\}$ denotes the trace of any matrix ${\bf A}$ (i.e. the sum of the diagonal terms) and the final equality is valid only for symmetric matrices such as the covariance matrix
The (\s) argument in equation (\ref{int1}) reminds us that the covariance matrix depends on the survey geometry chosen. The second approximate equality comes from the Cram\'er-Rao bound which states that the inverse of the Fisher matrix provides the best possible covariance matrix, see e.g. \cite{fisher2}. The equality is nearly exact when the likelihood is nearly Gaussian\footnote{Indeed, one can include almost-Gaussianity as a design criterion.}, defined by (e.g. \cite{fisher}):
\beq
F_{\mu\nu} = -\left\langle \frac{\partial^2 {\cal \ln L}}{\partial\theta_{\mu}\partial\theta_{\nu}}\right\rangle
\label{fisher1}
\eeq
where ${\cal L}$ is the likelihood. 

In eq. (\ref{int1}) $W^{\mu\nu}$ is a real, positive and symmetric matrix over ${\bf \Theta}$ which weights the importance of the various components of the covariance matrix, can be used to implement prior information, and can be chosen to optimally address specific questions, such as `is dark energy dynamical?' (see subsection \ref{dynamic}).

For example, if study shows that there are certain parameters, say $\theta_4$ and $\theta_6$, in the parameter space, which actually provide very little constraints on the underlying physics, it is clear that we should not give them equal weight in designing the survey. Instead we can down-weight their importance in the final optimisation process by making $W^{4\nu}$ and $W^{6\nu}$ smaller than the other components. In the limit where they are taken to zero, they will have no influence on the optimisation of the survey at all. 

The simplest choice is $W_{\mu\nu} = 1,~(\forall \mu,\nu)$; i.e. no priors, no $\thet$ dependence and all parameters weighted equally.
In this case and when the Fisher matrix is diagonal, eq. (\ref{int1}) reduces to:
\beq
I(s,\thet) = \sum_{\mu=1}^n \sigma_{\mu}^{-2}(s,\thet)
\label{tempi}
\eeq
i.e. the sum of the inverses of the expected variances of each of the dark energy parameters. Clearly we want to minimise the $\sigma_{\mu}$ and hence want to maximise expression (\ref{tempi}). 

However, instead of encoding prior information in ${\bf W}$ we can construct a slightly different $I$, which this time must be {\em minimised}:
\beq
I(s,\thet) = (({\bf F}(s,\thet ) + {\bf P})^{-1})_{\mu\nu} W^{\mu\nu}\,.
\label{prior}
\eeq
where ${\bf P}$ is the prior precision matrix and the Fisher Matrix is defined by (\ref{fisher1}).

Eq. (\ref{prior}) is useful since it automatically weights the importance of prior information correctly. If the new survey constraints (estimated by $F_{\mu\nu}(s,\thet)$) are much better than current data (which are summarised 
in ${\bf P}$) then the prior makes a negigible contribution to optimisation and visa versa. In this case it is natural to consider $W^{\mu\nu}$ independent of $\thet$. If we further believe that every parameter in ${\bf \Theta}$ is equally useful (a reasonable starting point if ${\bf \Theta}$ has been judiciously chosen), then we can simplify eq. (\ref{prior}) even further by choosing $W^{\mu\nu} = \delta^{\mu\nu}$ to arrive at\footnote{The Kronnecker delta satisfies $\delta^{\mu\nu} = 1$ if $\mu=\nu$ and 0 otherwise.} the following expression to be minimised:
\beq
I(s,\thet) = \mbox{tr}\{({\bf F} + {\bf P})^{-1}\}
\label{trace}
\eeq
where ${\bf F}$ depends both on $s$ and $\thet$. Minimisation of (\ref{trace})  corresponds to {\em A-optimality} in the optimal design literature \cite{L72}. However note that it does not give any importance to the off-diagonal terms in the covariance matrix $({\bf F}+{\bf P})^{-1}$.

In this special case $I$ reduces to 
\beq
I(s,\thet) = \sum_{\mu=1}^n \left(\sigma_{\mu, e}^{-2}(s,\thet ) + \sigma^{-2}_{\mu, p})\right)^{-1}
\label{special}
\eeq
where the $(e,p$) subscripts on the parameter variances $\sigma_{\mu}^{-2}$ denote the variances from the {\em expected (e)} survey data and from existing, {\em prior (p)}, data respectively. Again we see that if the new survey yields very small errors, as is expected for next-generation surveys, then $\sigma_{\mu,e}^{-2}$ is much larger than $\sigma_{\mu,p}^{-2}$ and dominates the maximisation process.

For complete generality let us now consider alternative definitions of $I$.

\subsection{Average error ellipse volume, D-optimality and maximum entropy}\label{detsec}

A slightly more elegant FoM can be defined, although it is rather more tricky to compute in nonlinear problems. In general we are trying to compute the survey configuration that minimises the error bars on our underlying parameters. Since these parameters typically exhibit degeneracies (particularly true of distance measurements) an obvious optimisation proceedure is to minimise the volume of the $n$-dimensional 1$\sigma$ error ellipse, averaged over ${\bf \Theta}$.

We can compute the volume of the error ellipse at any point in the parameter space ${\bf \Theta}$ via the square-root of the determinant of the covariance matrix. Equivalently one can consider the minimisation of the square of the volume, $V^2 \propto \mbox{det}(C_{\mu\nu})$ modulated by $w(\thet)$, a real-valued function that allows us to impose priors about what part of the parameter space ${\bf \Theta}$ is more important:
\beq
I(s,\thet) = \mbox{det}(C_{\mu\nu}) \tilde{w}(\thet)
\label{fomdet}
\eeq
Although this is a natural FoM, with an immediate geometric interpretation, it has the disadvantage of not being particularly easy to work with and, at least in this form, does not allow us to weight different parameters differently. 

Instead of trying to minimise (\ref{fomdet}) there are very good reasons to formulate this as a maximisation problem instead by considering:
\beq 
I(s,\thet) =  \tilde{w}(\thet) \log \mbox{det}({\bf F} + {\bf P}) 
\label{det}
\eeq
Maximising the integral of this expression is known in the optimal design literature as Bayes {\em D-Optimality}. 
We do not explicitly write the dependence on $(s,\thet)$ which we hope is obvious. Although maximising (\ref{det}) does not have the immediate geometric interpretation that minimising (\ref{fomdet}) has, it has a very powerful interpretation: namely it is the FoM which maximises the gain in {\em Shannon entropy} (hence the logarithm) or information. Equivalently it maximises the expected {\em Kullback-Leibler distance} in going from the prior to the posterior. In other words, it ensures that one gets the most information boost over what one already had in hand from the prior data alone. 

While D-optimality is clearly very powerful it, and the minimisation of error-ellipse volume expressed by eq. (\ref{fomdet}), will tend to favour survey configurations in which one of the principle axes of the ellipse is very small (the thinnest ellipse), and hence may not be ideal in many situations. 

For example, imagine that there exists a survey configuration such that the variance $\sigma^2_3$ becomes vanishingly small at every point in ${\bf \Theta}$. Minimising the FoM of eq. (\ref{fomdet}) would favour this survey configuration, but it might turn out that the corresponding parameter, $\theta_3$, which is measured with great accuracy, is actually of little use in constraining dark energy models. The survey would then provide wonderfully small error bars on an irrelevant parameter and large error bars on relevant parameters. Of course this problem can be easily resolved by choosing decorrelated variables $\thet$ that are physically useful. 

Which FoM is actually better for the needs of dark energy cosmology requires further research. We note that in the case where the covariance matrices are not diagonal, computing the inverse matrix is computationally intensive, so the relative CPU requirements of A- and D-optimality are not obvious.  

\subsection{Other figures of merit}

There are many other choices for $I$., known collectively as `alphabet'-optimality, including $C$-, $G-$ and $I$-optimlity and even finer schemes such as $D_{rm}$- and $G_{rm}$-optimality \cite{grm}, all optimising designs in different ways.
For example one could choose to find the thinnest possible error ellipses by maximising the biggest eigenvalue of the 
Fisher matrix. Alternatively one can 
minimise the maximum eigenvalue of the covariance matrix -- C-optimality -- which will favour small but circular error ellipses. 

This is also achieved by minimising the integral of:
\beq
I(s,\thet) = \frac{\mbox{Tr}\{({\bf F}+{\bf P})^2\}}{(\mbox{Tr} \{{\bf F}+{\bf P}\})^2} ~~~~~~~~~{\bf \odot -optimality}
\eeq
This expression is simply the ratio of the sum of the squares of all the entries of ${\bf F}+{\bf P}$ divided by the square of the trace and is maximised by diagonal Fisher matrices which in turn indicate decorrelated parameters and circular error ellipses.

It is even possible to construct uncountably many FoM using the fact that we can write the Fisher matrix (since it is square and symmetric) as the product ${\bf E}^T {\bf \Lambda E}$ where ${\bf \Lambda} = (\lambda_1,...\lambda_n)$ are the eigenvalues  and ${\bf E}$ is the set of eigenvectors of the Fisher matrix. Then we may define $I_p$ for $p \in [-\infty,\infty]$ as:
\bea
I_p(\thet,s) = && \mbox{max} ~\{\lambda_{\mu}\}~~~~~~\mbox{for}~~ p = \infty\nonumber\\
 && (\sum_{\mu} \lambda_{\mu}^p)^{1/p}~~~~~\mbox{for}~~ p \neq 0, \pm \infty \nonumber\\
 && (\Pi_{\mu} \lambda_{\mu})^{1/n} ~~~~~~\mbox{for}~~ p = 0\nonumber\\
 && \mbox{min} ~\{\lambda_{\mu}\} ~~~~~~~\mbox{for}~~ p = -\infty
\eea

\subsection{CPU constraints on the various FoM}

Splitting the Fisher matrix into two factors: derivatives which depend on $\thet$ but are independent of $s_i$, and errors which depend on $s_i$ (and sometimes are independent of $\thet$). Fortunately, the derivatives can be precomputed, saving time in the final optimisation. 

\begin{figure}
\begin{center}
\includegraphics[width=90mm]{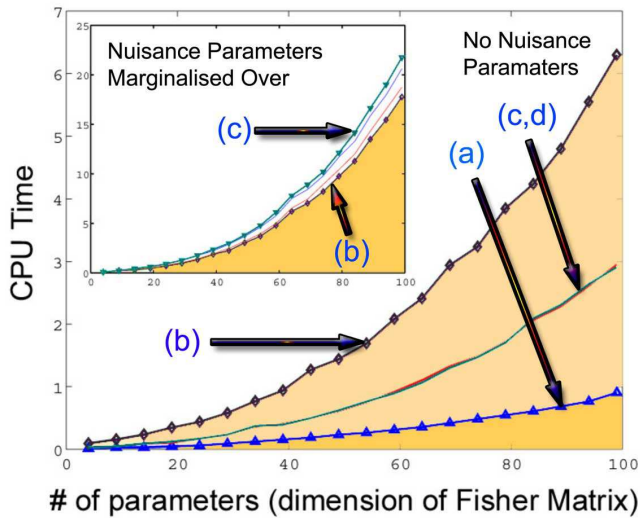} \\
\caption[survey]{CPU constraints on the various FoM for case without (main figure) and with (inset) nuisance parameters that must be marginalised over. The main figure (inset) shows CPU times for 2000 (1000) realisations respectively. The various curves are (a) Fisher-sum, (b) A-optimality, (c) Determinant and (d) D-optimality. Without nuisance parameters the Fisher-sum is significantly faster while A-optimality is the worst (since it is the only one that requires the inverse ${\bf F}^{-1}$ and since det(${\bf F}^{-1}$ = 1/det(${\bf F}$)). When nuisance parameters must be marginalised over the playing field is much more even and there is little CPU advantage to any of the FoM.} 
\label{cpu}
\end{center}
\end{figure}

Figure (\ref{cpu}) shows the CPU requirements for the various FoM we have considered as a function of the dimensionality of the Fisher matrix (number of parameters). The key observation is that nusiance parameters have a profound impact (see inset of figure), removing any CPU advantage of one FoM over another. If no nuisance parameters are included then CPU considerations will favour Fisher-sum optimality and strongly disfavour A-optimality. The inclusion of nuisance parameters (the realistic case) essentially removes this disparity because of the required matrix inversions which are common. 

We now move to the issue of how to actually compute the optimal survey configuration once an FoM has been selected.

\section{Optimising the Survey}\label{optim}

Survey optimisation occurs by computing the FoM for a range of survey geometries and simply selecting the geometry with the minimum/maximum FoM (depending on which of the candidate $I$'s was chosen).  In most situations (where the problem is nonlinear in the parameters $\thet$) analytical solutions for the FoM will not be available (although see below for special cases where there are) and instead we must compute it numerically. 

If the dimensionality of the survey configuration space, ${\bf s}$, is small this can be done using grid techniques or if the dimensionality of ${\bf s}$ is large (e.g. in the case where one has or order 100 redshift bins) one can use Monte Carlo Markov Chain (MCMC) techniques \cite{mcmc} with a standard engine (e.g. Metropolis-Hastings), with the FoM playing the role of the likelihood in standard implementations. Otherwise one can consider other extremisation methods \cite{NR}. 

\subsection{Finding the optimal niche in a competitive market}

Optimising the survey in the presence of existing or expected data sets from other experiments is easy to implement within IPSO. Consider a situation of interest in the current climate of survey design: optimising a survey given the expected SNAP and Planck data. 

To find the optimal niche given these or any set of `competing' surveys  we simply include, along with the prior information encoded in ${\bf P}$, the {\em expected} Fisher matrices for all the other surveys, denoted ${\cal F}^{SNAP}$, ${\cal F}^{Planck}$ etc... Hence one will now have: 
\beq
I(s,\theta) = {\cal O}({\bf F}(s) + {\bf P} + {\bf {\cal F}}^{SNAP} + {\bf {\cal F}}^{Planck} + ...)
\label{niche}
\eeq
where e.g. ${\cal O}({\bf A}) = \mbox{tr}({\bf A^{-1}})$ or $\log\mbox{det}({\bf A})$ depending on whether one chooses (\ref{trace}) or (\ref{det}) for $I$, or a corresponding generalisation for any of the FoM considered in section (\ref{fom}). Of these terms, only the first - ${\bf F}$ - actually depends on the candidate survey geometry, ${\bf s}$, although all terms (except ${\bf P}$) will in general also change depending on position in the parameter space ${\bf \Theta}$. The FoM will then include the information from SNAP, Planck and any other surveys included and extremising it will automatically select the survey geometry which provides the optimal niche given the other surveys. This is a systematic approach to 'orthogonalising' error ellipses for different experiments.  

In the case that all the matrices in eq. (\ref{niche}) are diagonal (or can be made diagonal by a single similarity transformation) then the A-optimal solution will be given simply by minimising the integral of $\mbox{tr}{\bf F}^{-1}$ since all the other terms ($\mbox{tr}{\bf P}^{-1}$ etc...) will simply add the same contribution independent of the candidate survey geometry (by linearity). In this case the weight matrix could be used to include prior information and expected results for other planned surveys. 

If on the other hand, at least one of the matrices in eq. (\ref{niche}) is not diagonal the inverse will not be a diagonal matrix in general and hence optimisation will depend on $\bf P$ etc... As a simple example consider the two parameter case ($\bf F, P$ etc... two-dimensional). The inverse is now inversely proportional to the determinant of the matrix sum in (\ref{niche}) and the mixing between the various matrices will remain after taking the trace.

In the case of D-optimality, we deal directly with the determinant of the sum in (\ref{niche}). Hence, even in the diagonal case the prior and SNAP/Planck etc... Fisher matrices make a crucial contribution to optimisation (for two or more parameters since otherwise the determinant is trivial). Consider again diagonal, two-dimensional $\bf F$ etc... so that $I$ becomes:
\bea
I(s,\thet) = &&\log (F_{11} + P_{11} + {\cal F}^{SNAP}_{11} + ...) +\nonumber \\
&&\log (F_{22} + P_{22} + {\cal F}^{SNAP}_{22} + ...)
\label{det2d}
\eea
All terms contribute to the final optimisation. We discuss the optimal solution for this case in the section on analytical optimisation. 

Returning briefly to the general expression (\ref{niche}), we can understand it geometrically in a simple way in terms of generating mock data. At every point in ${\bf \Theta}$ we generate mock data for each of the competing surveys, add it to the prior data (which might be all current SNIa data and is the same at every point in ${\bf \Theta}$ of course). This is done only once. Then, for each candidate survey ${\bf s}$ we generate mock data at each point in ${\bf \Theta}$ for the survey being optimised. This data changes with ${\bf s}$, the prior and competing data does not. One then computes the resulting FoM and extremises it.

\subsection{Dealing with nuisance parameters}

\begin{figure}
\begin{center}
\includegraphics[width=65mm]{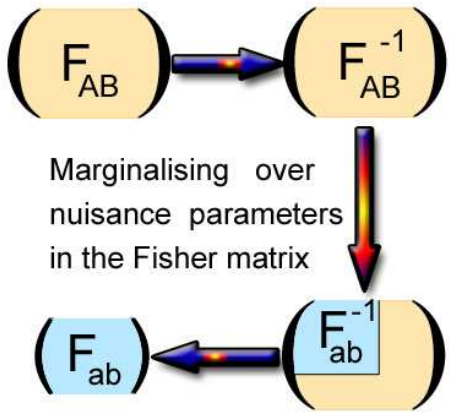} \\
\caption[survey]{IPSO integrates only over fundamental parameters $\theta_a$ while marginalising over the remaining nuisance parameters by extracting the appropriate sub-covariance matrix.}
\label{marg}
\end{center}
\end{figure}

Although we want to marginalise over the nuisance parameters $\theta_{\alpha}$ we do not (by definition of ``nuisance") want them to play an active role in the optimisation process. Hence we wish to extract the Fisher matrix $F_{ab}$ after marginalising over the nuisance parameters. This process is shown in Fig. (\ref{marg}) and is described in detail in \cite{se,fisher}. 

First, the full Fisher matrix, $F_{AB}$, is inverted. The sub-matrix is extracted corresponding to the rows and columns of the fundamental parameters, yielding $F^{-1}_{ab}$ which can be used directly in the FoM  (\ref{trace}) or inverted for use in (\ref{int1}) or (\ref{det}). The existence or not of nuisance parameters one must marginalise over has a significant bearing on computation constraints as shown in figure (\ref{cpu}).

\subsection{Addressing specific questions}\label{dynamic}

Survey designers may wish to optimise their survey in order to try to answer specific questions relevant for the day. This 
is particularly appropriate for experiments with short design and build lifetimes where one can be confident that the question of interest will not be answered before completion of the experiment. 

A good example is provided by one of the most pressing questions in dark energy research today, namely is the dark energy dynamical or is it a cosmological constant? It is plausible that by the time next-generation surveys reach final design state that there may still be no signal for dark energy dynamics (this will be the case if the true origin of acceleration is a cosmological constant). In which case we may be faced with the prospect of trying to detect dynamics (and hence rule out a cosmological constant) at the limit of the resolving power of even next-generation instruments. 

This question can be simply addressed: we want a survey to discriminate between a cosmological constant and dynamical models. Model-discimination favours use of D-optimality. We need a FoM which yields a survey which provides optimally small error bars around the parameter subspace corresponding to a cosmological constant ($w = -1$). This would require sacrificing accuracy in parameter regions far away from the cosmological constant which would be easily detected even with poor sensitivity in those regions.
Such a prescription can be easily implemented by choosing the weight matrix $W_{\mu\nu}(\thet)$ or weight-function $\tilde{w}$ in eq. (\ref{det}) appropriately. For example, if the cosmological constant corresponds to $\theta_{\mu} = 0$ for all $\mu$, then we could choose (for D-optimality)
\beq
\tilde{w}(\thet) = \exp\left(-\sum_{\mu,\nu}\beta_{\mu\nu}\theta^{\mu}\theta^{\nu}\right)
\label{gauss}
\eeq
For A-optimality one could choose ${\bf W} = \tilde{w} {\bf U}$ where every component of ${\bf U}$ is unity.

For $\beta_{\mu\nu}>0$ this will exponentially suppress the contribution to the survey design from regions of parameter space far from the cosmological constant parameter subspace, $\thet=0$. $\beta_{\mu\nu}$ measures the aggressiveness with which the suppression is implemented in each direction in ${\bf \Theta}$. But how should we choose $\beta_{\mu\nu}$? 

We want strong suppresion in the case where the prior indicates that any deviations from $\Lambda$CDM, if they exist, are small and if many models with strong dynamics (i.e. far from $\Lambda$CDM) are already ruled out, i.e. if prior variances on the variables are small which is equivalent to large entries in the prior precision matrix, ${\bf P}$. One choice which implements this idea is:
\beq
\beta_{\mu\nu} = \frac{P_{\mu\nu}}{(\theta^T_{\mu} - \theta^P_{\mu})^2}\,,
\label{wt}
\eeq
where $\theta^T_{\mu}$ represents the parameter values of the {\em target} model we are trying to rule out (in this case  $\Lambda$CDM and $\theta^T_{\mu}=0$).  $\theta^P_{\mu}$ on the other hand are the maximum likelihood estimators of the $\thet$ using only the {\em prior} data. In practise this would benefit from softening to keep the components of $\beta_{\mu\nu}$ finite in chance cases where one or more of the target parameters coincide to high accuracy with the prior best estimates of those same variables.

What this choice does is the following: if the best-fit to prior data is close to the target model and if the prior parameter variances are small, then the suppression will be strong, reflecting the need to aggressively optimise the error bars near the target model. If, on the other hand, the prior-data best-fit is far from the target, or the data is very poor so the variances are large, then the resulting suppression is weak, i.e. small $\beta_{\mu\nu}$. The expression (\ref{wt}) is appropriate whenever the target model corresponds to a single point in ${\bf \Theta}$. 

In the case where one is trying to descriminate between two classes of models corresponding to sub-spaces, ${\bf \Gamma_{1,2}} \in {\bf \Theta}$, which both have non-zero volume in ${\bf \Theta}$, ($\mbox{Vol}_{\bf \Theta}({\bf \Gamma_{1,2}}) > 0$) a different suppression expression is needed since a single target point nolonger exists. In this case we propose to replace the denominator of (\ref{wt}) with an estimator of the minimum distance between the two sub-spaces. If the sub-spaces are far apart it will be much easier to differentiate them than if they are close to each other. Therefore we propose:
\beq
\beta_{\mu\nu} = \frac{P_{\mu\nu}}{d(\Gamma_1,\Gamma_2)}\,,
\label{wt2}
\eeq
where 
\beq
d(\Gamma_1,\Gamma_2) = \mbox{min}\left\{d(\vec{\theta}_{1},\vec{\theta}_{2})~|~ \vec{\theta}_{1} \in {\bf \Gamma_1}, \vec{\theta}_{2} \in {\bf \Gamma_2} \right\}
\eeq
and where $d(\vec{\theta}_{1},\vec{\theta}_{2})$ gives the distance between any two points $(\vec{\theta}_{1},\vec{\theta}_{2})$ in the natural metric of ${\bf \Theta}$ (which will typically be a flat, Euclidean space). As an interesting aside, in the case where either subspace is disconnected one can still use this expression although there may be more optimal choices. As an example of disconnected subspaces consider the kink parametrisation of dark energy \cite{longDE,params}. There models indistinguishable from $\Lambda$CDM correspond to two disconnected subspaces; namely $(w_0=-1,w_m=-1)$ with $(\Delta, a_t)$ arbitrary on the one hand and $(w_0=-1, a_t<10^{-4})$ and $w_m$ arbitrary (with $\Delta$ sufficiently large) on the other.  

For any of these prescriptions however, the extreme limit $\beta_{\mu\nu} \rightarrow \infty$ yields a matrix of delta-functions for $W_{\mu\nu}$. In this case the integration over ${\bf \Theta}$ becomes trivial and optimising the survey collapses to finding the smallest error bars at the single chosen point, $\theta^T_{\mu}$, in the parameter space. Of course, building a survey to best address a specific question reduces the cross-section for serendipitous discovery. As an example, consider the issue of dark energy dynamics once again. 

Concentrating around the cosmological constant, as implied by eq. (\ref{gauss}), may (depending on the parameter space ${\bf \Theta}$) maximise sensitivity at intermediate redshifts ($<1$) in an effort to detect deviations from the predictions of $\Lambda$CDM. As a result, the survey design may have no sensitivity at very high redshift $z>3$ where dark energy is irrelevant in the $\Lambda$CDM model and hence the ability to make serendipitous discoveries (such as a sudden input of radiation or an oscillating dark energy equation of state) is weak. Having said this, a sufficiently good choice of parameter space, ${\bf \Theta}$, will allow for this possibility, since serendipitous discovery at high-z would automatically imply dynamics. 

\subsection{Implementing survey constraints}

Each candidate survey geometry, indexed by ${\bf s}$, will differ from other candidates in one of two conceptually different ways, which we label {\em hard} and {\em soft} differences. Hard differences correspond to discontinuous, radical, changes in survey structure. Tyically hard differences will be unique to each experiment and correspond to proposals to supplement the fiducial survey with a fundamentally different component. The possibility of including a weak lensing survey in the SNAP mission strategy is a good example of a hard survey change\cite{snaplens}.

Typical next-generation dark energy experiments will spend considerable time (1-5 years) surveying a considerable fraction of the sky with high resolution and obtaining both spectra and photometry data at multiple frequencies. 
As a result many opportunities exist for overlap between the various dark energy tests. For example, the proposed Dark Energy Survey \cite{des} would obtain constraints on dark energy from cluster counts, weak lensing, type Ia supernovae, baryon oscillations and extraction of the ISW effect through cross-correlation with Planck. The main question is whether spending time on the extra component justifies the corresponding loss of constraints from the main part of the survey (in the case of SNAP, detection of SNIa).

On the other hand, soft differences in survey geometry correspond to smooth changes in the parameters underlying the  survey. For example, at fixed total observing time, what is the optimal observing time for each redshift bins? Or, what is the optimal split between survey area and survey depth? 

If hard differences are being considered then the output of the optimisation must answer whether they are justified and if so, what the optimal resulting soft configuration should be \footnote{In reality, treating hard differences fully is  nontrivial because the resulting constraints will depend sensitively on engineering, manufacturing and design ingenuity in integrating the hard changes with the fiducial survey.} In a follow-up paper we will consider optimisation of the KAOS survey \cite{kaos} given expected constraints from the SNAP satellite. 

For linear problems (see e.g. section \ref{volume}) it is possible to perform the optimisatation analytically. In most realistic cases, however, this will not be possible. If there are many different candidate survey geometries then it is likely that MCMC methods will be the prefered optimisation technique. The disadvantage of the MCMC method is that it is not naturally suited to deal with both hard and soft differences, but only soft differences where the space of survey geometries is connected. In the case of hard differences it may not always be possible to make the space of survey geometries connected. In such a situation grid or mixed grid-MCMC methods may have to be used. 

In most cases, however, hard differences can be simply accounted for by considering the combined space containing the survey parameters for each of the components of the survey. For example, consider the case of SNAP and whether it should undertake a weak lensing survey (with `soft' parameters $\{\delta_i\}$) in addition to the baseline SNIa survey (with `soft' parameters $\{\gamma_i\}$). MCMC techniques could be used on the combined set $\{\delta_i,\beta_i\}$ as long as 
the survey geometry with no weak-lensing occupies a finite volume in the full space (i.e. does not just correspond to a measure zero subspace), otherwise the random sampling inherent in MCMC will never select the single survey geometry. 
Clearly, implementation will depend on the specific survey under consideration.

\section{Examples of optimisation}\label{egs}

Later we will consider how to construct the optimum error bars in a set of $n$ redshift bins located at {\em fixed} redshifts $z_i, ~i=1...n$ with variable error bars, $\epsilon_i$. First we consider the opposite problem: that of one-point optimisation.

\subsection{Where is the golden point? One-point optimisation}

If we are allowed to make one observation (with fixed error bar independent of redshift), then at what redshift should we place it? We will call this the golden point problem.  Unlike true optimisation there is no constraint to implement (since we assume the error bar is independent of $z$). Hence the problem is much simpler and will allow us to get a feeling for some of the issues involved in optimisation. 

A simple analytical solution exists in the case where we also only consider a one dimensional parameter space, ${\bf \Theta}$ since then the various FoM become trivial. Further, in the case that the Fisher `matrix', $F$, is independent of $\theta$ then one can immediately see that the three proposed optimisations: A-optimality, maximising $\mbox{det}(F)$ and D-optimality, all become equivalent. This follows since minimising $F^{-1}$ is clearly the same as maximising $F=\mbox{det}(F)$ or $\log \mbox{det}(F)$ since the logarithm is a monotonic function. On the other hand, as soon as $F = F(\theta)$, the three optimum solutions will no longer coincide in general. Clearly the golden solution will also depend on what quantity is being measured. 

We will consider two examples of one-point optimisation (see also \cite{HT} for a similar discussion but without the integration over ${\bf \Theta}$). The first is to determine the $H(z)$-golden point in a flat $\Lambda$CDM universe where the parameter is $\theta = \Omega_m$, the matter density today and $\Omega_{\Lambda}=1-\Omega_m$. We will consider for simplicity a measurement of the Hubble constant. Consideration of $d_A(z)$ or number counts is analagous.  The Fisher point (the only component of the Fisher matrix), $F$ is then written in terms of $E(z) \equiv H(z)/H_0$ as:
\beq
F \propto \left(\frac{\partial E(z)}{\partial \Omega_m}\right)^2 \propto \frac{((1+z)^3 - 1)^2}{E^{2}(z)} 
\label{fp}
\eeq
where we have dropped normalisation constants which will not affect the optimisation (such as $H_0$ and $\epsilon$, the redshift-independent error-bar). The game is then to find the value of $z$ which extremises the corresponding FoM. 
The three optimisations then correspond to minimising $\int F^{-1} d\Omega_m $ or maximising either $\int F d\Omega_m$ or $\int \mbox{log}(F) d\Omega_m$ (A-optimality, determinant and D-optimality respectively).

\begin{figure}
\begin{center}
\includegraphics[width=80mm]{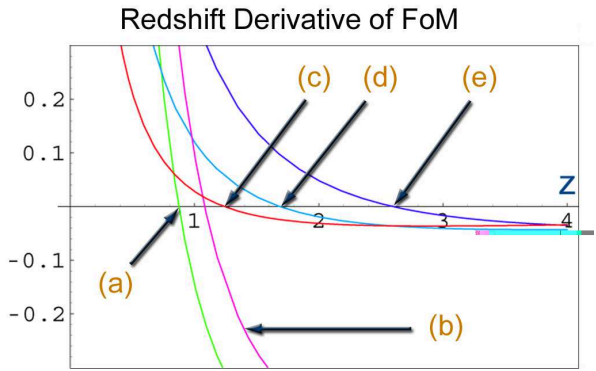} \\
\caption[survey]{{\bf Priors and parameter-space averaging affects the golden point}. The one-point D-optimum redshift
for Hubble constant measurements is given by the zeros of the curves (a)-(e) for a constant $w$ with different priors on the parameter space $\theta = \{w\}$. The various curves correspond to top-hat priors with upper and lower limits on $w$, i.e. $((w_{lower},w_{upper})$ of: (a) $(-1.5,-0.8)$,  (b) $(-1.5,-0.6)$, (c) $\Lambda$CDM only, (d) $(-1,-0.8)$ and (e) $(-1,-0.6)$. The optimum redshift varies from $z_{opti} = 0.86$ (a) to $z_{opti} = 2.6$ depending on the prior assumed. Allowing larger values of $w$ increases the importance of the dark energy at higher redshift and naturally shifts the golden point to higher redshift. Conversely, for more negative lower-bound forces the golden point to smaller redshifts and makes the change in the optimum sharper.}
\label{1pt}
\end{center}
\end{figure}

Looking at eq. (\ref{fp}) we immediately see that assymptotically $F \rightarrow (1+z)^3$ for large $z$ since the derivative of $E$ w.r.t. $\Omega_m$ is monotonically increasing and hence $F$ is maximised by simply going to the largest available redshift. $\log(F)$ shares the same property so D-optimality makes the same prediction. 

This same conclusion is reached if we consider a different example. Let us fix $\Omega_m$ and choose $w(z)$ so that $w(z) =-1$ for $z<z_*$ and $w(z)=0$ for $z \geq 0$. This is a crude model for a very rapid transition of the type that is actually a rather good fit to current SNIa data \cite{params}. What is the golden redshift once we average over $z_*$, the only parameter in the problem? A brief calculation confirms the obvious: $z_*$ is determined by the expansion rate which is best measured at high redshift. Hence, perhaps counter-intuitively, the best way to detect a sudden transition through  $H$ is by a measurement at high redshift (assuming redshift-independent error bars).  

Now let us consider an example in which the golden point is not the maximum  redshift available to the survey. Again consider a single Hubble constant measurement but this time let us consider $\theta=\{w\}$, i.e. consider a constant equation of state parameter for the dark energy. We will again assume a flat universe and this time with a $70/30$ split between the energy density of dark energy and matter today. None of these assumptions is crucial and in a full analysis one would include these parameters in ${\bf \Theta}$ to be integrated over in the optimisation.

For a general parametrisation of the dark energy equation of state we have
\beq
E^2 = (\Omega_m (1+z)^3 + (1-\Omega_m) f(z,\theta_{\mu})
\eeq
where $f$ controls the time evolution of the dark energy:
\beq
f(z,\theta_{\mu}) = \exp(3\int\frac{1+w(\theta_{\mu})}{1+z}dz)
\eeq
In this case, we have $f=(1+z)^{3(1+z)}$. Again the Fisher matrix is a single point, but this time the relevant derivatives are not necessarily monotonic. For a general single parameter expansion of $f$ we have:
\beq
F \propto \frac{f^2}{E^2}\left(\int \frac{1}{1+z}\frac{\partial w}{\partial \theta}dz\right)^2
\label{oneptw}
\eeq
$f/E^2$ is the ratio $\rho_{DE}/\rho_{tot} = \Omega_{DE}$. Since in most standard dark energy models $\Omega_{DE} \rightarrow 0$ with increasing $z$ we see that measuring dark energy parameters will favour observations at lower redshift. 
In the case of constant $w$ the redshift-integral collapses to $\log(1+z)$. 

To compute the golden point we need to compute the FoM and then solve the equation $\partial FoM/\partial z = 0$. This is done most easily numerically. Figure (\ref{1pt}) shows $\partial FoM/\partial z$ as a function of $z$ and the golden point corresponds to the zero of this function. In the figure we consider only D-optimality defined by eq. (\ref{det}) and neglect the prior precision matrix ${\bf P}$. 

However, in computing the FoM we must integrate over $w$. But what range of $w$ is reasonable to consider? The weak energy condition imposes $w\geq -1$ and $w>1$ seems unphysical. Such priors can be implemented through the weight function $\tilde{w}(\theta)$ in eq. (\ref{det}). What effect does imposing such theoretical priors have on the golden point? 

Figure (\ref{1pt}) shows several different curves which differ only in the choice of $\tilde{w}(\theta)$. In each case we choose it to be a top-hat function which is zero outside of the range $(w_{lower},w_{upper})$. As we increase $w_{upper}$ we include solutions into the optimisation which are more and more dynamically important at high redshift (since $\rho_{DE} \propto (1+z)^{3(1+w)}$) and hence the golden point moves to higher redshift. Conversely, if we decrease $w_{lower}$ the dark energy becomes less and less apparent at high redshift and the golden point rapidly shrinks towards zero. 

This simple example illustrates two key points: (1) Not integrating over the parameter space but simply taking a fiducial model such as $\Lambda$CDM (corresponding to curve (c) in fig. (\ref{1pt}) looses a great deal of the potential to optimise a survey. (2) Given the infinite dimensional space of dark energy models, optimisation cannot be done without specifying the class of dark energy models one wants to detect.

As a more sophisticated example, let us parametrise the dark energy energy density in terms of $f(z)\equiv \rho_{DE}(z)\rho_{DE}(0)$ by considering $n$ independent redshift bins centered for convenience at $z_j = j\Delta$ with
\beq
f(z_k) = 1 + \sum_{j=1}^k \theta_j
\label{fbin}
\eeq
where $\theta_k \equiv f(z_k)-f(z_{k-1})$ control the change in $f$ between bins. The WEC ($w \geq -1$) corresponds to $\theta_j \geq 0$ for all $j$. Let us consider the golden-point problem with the slight simplification that we only consider putting our observation at one of the redshifts $z_j$. Note that the parameters are perfectly correlated: a change in any of the $\theta_j$ can be compensated for by an opposite change in any of the other parameters. A short calculation shows that the Fisher matrix for this system is very special: 
\beq
F_{ab}(z_i) \propto E^{-2}(z_i) M_{ab}
\eeq
where $M_{ab} = 1(0)$ if $max(a,b) \leq (>) i$. In other words, the Fisher matrix vanishes beyond the bin at which the observation is being made. As a result ${\bf F}$ is singular, $\mbox{det}(\bf F)$, reflecting the perfect correlation of the parameters (the error ellipses are lines with zero volume). In fact this is a rather general property of one-point optimisation since one is constraining the function at a single point. This perfect correlation disappears when we have two or more measurements since one begins to test the shape of the function.

\begin{figure}
\begin{center}
\includegraphics[width=85mm]{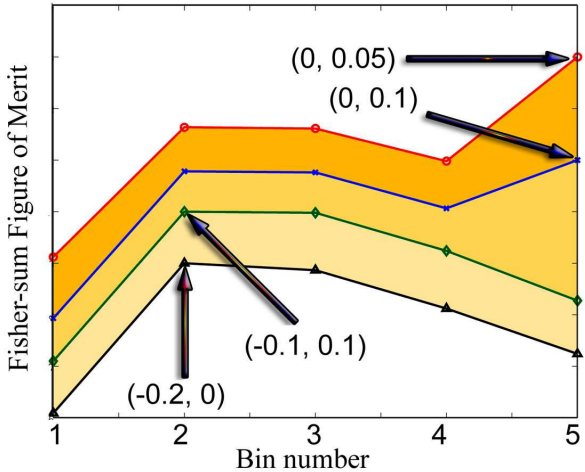} \\
\caption[survey]{Four different optimisations differing only by the values of $(\theta_{min},\theta_{max})$. Negative $\theta_{min}$ pushes the optimum to lower redshift. However, due to the non-trivial integration structure the results are not obvious and there are multiple extrema. }
\label{stepfig}
\end{center}
\end{figure}

Because of this degeneracy we cannot apply A-optimality, D-optimality nor determinant optimality. However, Fisher-sum optimality will work, eq. (\ref{int1}), where we take all the entries of $W_{ab}$to be equal top-hat functions, equal to unity for $\theta_i \in [\theta_{min},\theta_{max}]$ and zero outside. This allows us to impose prior constraints (such as implementing the WEC which implies $\theta_{min} \geq 0$).  

The FoM then becomes:
\bea
FoM(z_i) &\propto& \int_{\theta_{min}}^{\theta_{max}}  \sum_{a,b}^n F_{ab}(\theta_j)d\theta_1...d\theta_n \nonumber \\
&&= i^2 V^{n-i} \times \int_{\theta_{min}}^{\theta_{max}} \frac{d\theta_1...d\theta_i}{E^{2}(z_i, \theta_j )}
\eea
where $V\equiv (\theta_{max}-\theta_{min})$ and where we are trying to find the optimum value of $i$ which determines the optimum redshift. 

Figure (\ref{stepfig}) shows the resulting FoM and optimal bin numbers (marked with arrows) as a function of $i$ and show the strong dependence on the choice of $\theta_{min}$ and $\theta_{max}$ which are inputs from the survey designer. Again this highlights the need for precise design requirements.

\subsection{Analytical optimisation and linearity}\label{volume}

There are certain cases where the optimisation can be performed analytically. The standard example is the case where the covariance matrix and weight matrix $W_{\mu\nu}$ are independent of the underlying parameters $\thet$, in which case the integration of any of the choices for $I$ is trivial, yielding a weighted volume factor that is the same for all survey geometries and hence is irrelevant to optimisation. In this case, the problem reduces to one amenable to analytical solution using, e.g. Lagrange multipliers. 

Consider A-optimality, given by eq. (\ref{trace}), in the limit in which ${\bf F}$ and ${\bf P}$ are diagonal and the problem is linear so the Fisher matrix does not depend on the $\thet$. Under these approximations ${\bf P}$ makes no contribution to the optimisation and we have (c.f. eq. \ref{special})
\bea
&&\sigma_{\mu}^2 = \left[\sum^{N_p}_{i=1} \frac{1}{\epsilon_i^2}
\left(\frac{{\partial X_i}}{\partial \thet}\right)^2\right]^{-1} \\ \nonumber
&&FoM \propto \sum^{n}_{\mu = 1} \sigma^{2}_{\mu}~~~~
\label{1sig}
\eea
where $N_p$ is the number of data points or redshift bins in the survey  and $X_i = X(z_i)$ is the underlying variable of interest (such as luminosity or angular diameter distance). In general $N_p$ may well vary depending on the survey geometry, but here we consider it fixed.

We are interested in finding the optimal experimental error bars we should achieve for each redshift bin, $\epsilon_i$, which as a group maximize the FoM. The $\epsilon_i$ are the direct output from each survey geometry, $s$. While in any specific application it is the (soft) survey parameters that we would vary and optimise over, for generality we will find the optimimum $\epsilon_i$ distribution. 
\begin{figure}
\begin{center}
\includegraphics[width=70mm]{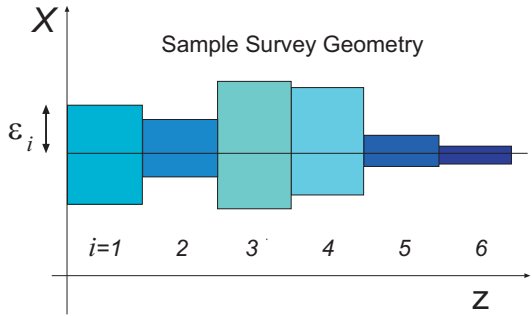} \\
\caption[survey]{Schematic illustration of a survey geometry. For a given observable quantity of interest, $X$ (such as luminosity distance), the $1\sigma$ error bars, $\epsilon_i$, in each of several redshift bins (here $i=1...6$) are free  to vary subject only to a constraint such as eq. (\ref{constraint}).}
\label{sigma_fig}
\end{center}
\end{figure}
Of course, without any further constraint, any reasonable FoM will be optimised by choosing $\epsilon_i^2 = 0,~~ \forall i$. Clearly we need to impose constraints such as cosmic variance, shot noise, finite resolution, systematic errors, finite observing time etc... As a model of such errors we optimise the FoM subject to the constraint:
\beq
f(\epsilon_i) = \sum_{i=1}^{N_p}\left(\frac{\alpha_i}{\epsilon_i^n} - \frac{1}{\epsilon_*}\right) \leq 0\,.
\label{constraint}
\eeq
The optimal solution saturates the bound. This is a simple encoding of the physical constraint that some weighted sum of the errors must be larger than a minimum, $\epsilon_*$ and ensures that the error in each bin is bounded from 
below by $\epsilon_j \geq (\epsilon_*/\alpha_j)^{1/n}$, so that no one error bar can be made arbitrarily small (as would be 
allowed if we considered a linear sum of the $\epsilon_i$ instead). 

Here the $\alpha_i$ characterise the efficiency of observing in the $ith$ bin while $n$ quantifies the nonlinearity of the constraint. It is typical that error bars at high-z are worse than those at lower redshift given equal resources 
(e.g. high-z objects are typically fainter and require longer observing time to get a spectrum or a good light curve in the case of SNIa). Hence in this example we expect that the $\alpha_i$ would be an increasing 
function as $i$ (and $z$) increases expressing the increased cost of obtaining constraints at high redshift. An effective dependence of $\alpha_i \propto (1+z)^6$ has been suggested in the literature for follow-up of SNIa\cite{SS}.

A naive construction of a typical constraint is as follows: assume the total observing time, $T$, for the survey is fixed and we are able to devide up this time between target objects in various redshift bins, appropriate in the case where target position and redshift are already known. The resulting error bars will typically scale as the square root of the observing time (or the number of objects) in each bin and hence $n=2$. However there is no solution for $n=2$\footnote{For $n=2$ the minimum occurs on the boundary of the allowed region but does not correspond to a point where the first derivatives vanish. For $n>2$ such a local minimum does appear within the physical region allowed by the constraint.} while for $n<2$ we have only a minimum (i.e. the worst possible solution). For $n>2$ we have a maximum and we therefore focus on this case. 

We can now find the optimal distribution of error bars analytically using Lagrange multipliers. Consider the function $y = FoM + \lambda f(\epsilon_i)$. The extremum of $y$ also corresponds to the extremum of the $FoM$ as long as we impose $\partial y/\partial \lambda = 0$ which enforces the constraint $f(\epsilon_i)=0$. Solving the resulting set of equations found from $\partial y/\partial \epsilon_i = 0$ gives the minimum FoM and optimal error bars (for $n > 2$). {\em For a single parameter ($\theta_{\mu}$) we have:}
\bea
\epsilon^{n-2}_i &=& \alpha_i\left(\frac{{\partial X_i}}{\partial \thet}\right)^{-2} \times Y \\ \nonumber
Y &\equiv& \epsilon_*^{\frac{n-2}{n}}\left(\sum^{N_p}_{i=1}\alpha_i \left(\sqrt{\alpha_i}\frac{\partial X_i}{\partial \thet}\right)^{\frac{2n}{2-n}}\right)^{\frac{n-2}{n}}
\label{optimal} 
\eea
$Y$ is simply a normalisation that sets the overall scale of the error bars but is otherwise irrelevant.
We remind the reader that $n$ is determined by the constraint (\ref{constraint}). The key point of this simplified solution is that it gives an intuitively reasonable answer: one should spend most of ones time getting small error bars (i.e. observing the most) where efficiency is high (small $\alpha_i$) and where the resulting constraints on the parameters, $\thet$, are large, {\em viz:} where the $\frac{\partial X_i}{\partial \thet}$are large. 

This solution is applicable (at least in the Fisher Matrix approximation) whenever the underlying quantity of interest (such as the distance) is linear in each parameter and we are only interested in soft changes to the survey (i.e. $n$ and the $\alpha_i$ are always the same). 

In the realistic case where we are optimising w.r.t. multiple parameters ($P$ is two or higher dimensional) equation (\ref{optimal}) is modified in a natural way (in each bin $i$ we now have a sum over the $n$ derivatives of $X$ wrt. the parameters $\thet$). We do not give the formula explicitely since in most practical applications a numerical optimisation will be easiest and most efficient to implement.

What does our alternative FoM, that is D-optimality, formulated in terms of the determinant give? Here we include the prior precision matrix and expected Fisher matrix from SNAP (see eq.\ref{niche}). Consider the simplest case where the covariance matrices are diagonal with two parameters and independent of $\thet$. In this case optimising the FoM is equivalent to optimising just the determinant without the logarithms in (\ref{det2d}) or (\ref{det}), since the logarithm is a monotonic function. Solving the resulting system for two parameters subject to the same constraint, eq. (\ref{constraint}) (which defines $n$), gives:
\bea
\epsilon^{n-2}_i &&= \frac{\alpha_i}{\left(\frac{{\partial X_{1}}}{\partial \thet}\right)_i^{2}\left[A_{22}\right]  +\left(\frac{{\partial X_{2}}}{\partial \thet}\right)_i^{2} \left[A_{11}\right]} \times Y \nonumber \\
A_{\mu\nu} &&\equiv F_{\mu\nu} +  P_{\mu\nu} + {\cal F}^{SNAP}_{\mu\nu} + ...
\label{detop}
\eea
where $Y$ is simply a constant normalisation as before. The generalisation to more than two parameters is trivial.
The resulting optimum $\eps$ have the same basic structure as (\ref{optimal}) - namely they are proportional to the $\alpha_i$ and inversely proportional to a weighted sum of the derivatives of $X$ with respect to the parameters $\theta_1,\theta_2$. 

The main difference from diagonal Fisher matrices implementing A-optimality is that now the prior precision matrix and expected errors from SNAP, Planck etc... play an important part in determining optimal strategy. Naively one might expect that, everything else being equal ($\alpha_i$, X-derivatives etc...), one should look where prior and competing experiment constraints are bad, i.e. in the `data-desert'. Instead, at least in this case, it tells us that we should concentrate on the regions where constraints are already good and where competing experiments will focus (since there $A_{\mu\nu}$ is largest). 

As an interesting example, consider a baryon oscillation survey such as KAOS \cite{kaos,blake,se,linder}, which would yield excellent constraints both on the Hubble `constant' and the angular-diameter distance as a function of redshift out to $z=3$ and beyond. Which part of the redshift range should KAOS focus? Considering the angular diameter distance case first and the fact that SNAP and the currently highest known redshift SNIa go only to $z=1.7$, it is clear that the region beyond $z=1.7$ will have $A_{\mu\nu} = F_{\mu\nu}$ and so optimal error bars should be larger there. Indeed, there may be a good argument for not going beyond the optical, viz $z=1.3$. 

The Hubble constant constraints are very different since there are essentially no current constraints and SNAP will give no direct constraints on it. A proper optimisation of KAOS would have to include both data sets and is left to the future.

\subsubsection{Small off-diagonal terms don't matter for D-optimality}

Most analytical work on covariance matrices (such as that above) assumes diagonality, i.e. that the covariances between parameters vanishe. However, we can derive an interesting result relevant to D-optimality when there are small off-diagonal terms, {\em viz}
\beq
{\bf F} = {\bf F_0}(\mbox{\bf I} + \epsilon {\bf F_1})~,~~\epsilon \ll 1
\label{expand}
\eeq
where here $\mbox{\bf I}$ is the unit matrix, ${\bf F_0}$ is diagonal and ${\bf F_1}$ carries only off-diagonal terms. Then we can solve perturbatively (using the general relation $\ln \mbox{det}~{\bf C} = \mbox{tr} \ln {\bf C}$ for any matrix ${\bf C}$) to find
\beq
\mbox{det}({\bf F}) = \mbox{det}({\bf F_0})[1 + \epsilon ~\mbox{tr}({\bf F_1}) + {\cal O}(\epsilon^2)]
\label{smallcov}
\eeq
But since our splitting into ${\bf F_0}$ and ${\bf F_1}$ was done so that all the diagonal entries of ${\bf F_1}$ vanished we have:
\beq
\mbox{det}({\bf F}) = \mbox{det}({\bf F_0}) + {\cal O}(\epsilon^2)
\eeq
Hence, the assumption of diagonality is a good one for D-optimality as long as the off-diagonal terms are reasonably small. D-optimal solutions are stable under small non-diagonal perturbations of the covariance/Fisher matrices.

Unfortunately the same does not hold for A-optimality since the inverse of eq. (\ref{smallcov}) will contain a term linear in $\epsilon$. Hence we can expect small non-diagonal terms to cause a bigger shift on A-optimal solutions than D-optimal ones. However since ${\bf F_0}$ is diagonal one can easily and systematically compute corrections to the A-optimality FoM to any order in $\epsilon$. Systematically correcting optimal solutions for such perturbations is an interesting issue for further study.

\subsubsection{A simple example}

We end with a simple but relevant example where analytical optimisation applies. Consider the luminosity distance, $X=d_L(z)$, Taylor expanded in some general basis functions $Y_i(z)$ which could be nonlinear in $z$ (such as $Y_i \equiv (1-a)^i = \frac{z^i}{(1+z)^i}$): 
\beq
d_L(z) = d_0\left[1 + \theta_1 Y_1(z) + \frac{\theta_2}{2} Y_2(z) + ...\right]
\label{distance}
\eeq
The key point is that this formula is linear in the $\thet$, and hence the problem reduces to that in eq. (\ref{1sig}). On the other hand, had we chosen to parametrise the dark energy equation of state $w(z)$ or energy density $\rho(z)$, then the dependence of $d_L(z)$ on the $\theta_{\mu}$ would be nonlinear in $\thet$ and the covariance matrix would need to be integrated numerically over ${\bf \Theta}$.

This raises the question - which fundamental quantity is the best to use in designing a survey? Clearly avoiding integration over the parameter space is useful from a computational view, especially if the optimisation can then be done in large part analytically. This can of course be achieved by using a parameter space which linearly parametrises the quantity of interest, as in the example above, although the connection to fundamental physics will not be clear (hence one is not clear that one is optimising with respect to fundamental physics). This interesting issue is left to the future.

\section{Conclusions and future work}

Mutual optimisation of surveys is important if we are to maximise the knowledge extracted from next-generation experiments about cosmology and the true nature of dark energy in particular. Given the profound impact that understanding dark energy will have on fundamental physics it is a worthy endeavour. In this paper we propose a framework for survey design and optimisation which is model-independent (does not assume any specific underlying dark energy or cosmological model), maximises serendipitous discoveries, is flexible (the user can define their own parameter space to be integrated over) and allows optimisation of a survey given other `competing' surveys such as SNAP or Planck. In this way the best niche for a survey can be found precisely and quickly. 

This framework - Integrated Parameter Space Optimisation (IPSO) - is set-up within the context of Bayesian optimal design and naturally allows for best use of prior information in the design phase. The basic idea of IPSO is to extremise a {\em Figure of Merit} (FoM) - found by integrating the 1$\sigma$ errors for that geometry over the entire dark energy parameter space ${\bf \Theta}$ - in the set of candidate survey geometries under consideration. 

We have considered three main FoM which correspond to minimising the average sum of parameter variances, minimising the average error ellipse volume and maximising the gain in Shannon information in going from the prior to the posterior. The first and last FoM correspond to A-optimal and D-optimal solutions in Bayesian optimal design literature when the weight matrix takes on a special form. We have further shown that D-optimal and the minimum average volume solutions, obtained assuming diagonal covariance matrices, are stable to small off-diagonal perturbations, while A-optimal solutions are not. 

The parameter space integration allows these FoM to be model-independent and to be sensitive to the entire gambit of dark energy models while mutual optimisation with respect to other surveys is achieved trivially by including their effects in the forecasting of the $1\sigma$ errors. Optimisation then automatically chooses a survey configuration to ``orthogonalise" the resulting sets of constraints and find the best niche for the survey being designed in the crowded market place.

Some of the questions that will need to be addressed in future work is which FoM to use for dark energy studies, how to optimally choose the parameter space ${\bf \Theta}$ and the weight matrix $W_{\mu\nu}$, especially to best answer specific questions, such as whether dark energy is dynamical or not. The latter cannot be decided upon until ${\bf \Theta}$ is fixed. But even when this is done, choice of $W_{\mu\nu}$ is highly non-trivial if one ventures away from unity. Also of interest is the numerical implementation of IPSO and how FoM change under a smooth change of ${\bf \Theta}$. Detailed study should be undertaken to determine these interesting issues. A conservative first step is to use either the A-optimal or D-optimal expressions, equations (\ref{trace}) and (\ref{det}) respectively. 

Nevertheless, the fact that optimised survey design is now a question worth addressing seriously reflects the rapid gain in maturity of modern observational cosmology and illustrates the coming profound shift to over-determined science where each of the inputs to cosmology is strongly constrained from multiple vantage points. The golden age of cosmology will be a show worth keeping ones eyes open for.

\section*{Acknowledgements}

I thank Takairo Tanaka for many excellent suggestions and discussions and Martin Kunz who stimulated me to consider the link to Bayesian design methods. I also warmly thank Chris Blake, Josh Bryer, Martin Kunz, Bob Nichol and other members of the KAOS $w(z)$ science feasibility study team for interesting discussions which stimulated my initial interest in this topic. This research was supported by a Royal Society/JSPS fellowship.

\end{document}